# Smart meta-superconductor MgB$_2$ constructed by inhomogeneous phase of luminescent nanocomposite

Yongbo Li, Honggang Chen, Mingzhong Wang, Longxuan Xu and Xiaopeng Zhao*

Smart Materials Laboratory, Department of Applied Physics, Northwestern Polytechnical University, Xi'an 710072, China; lybo2010@foxmail.com (Y.L.); 2017100698@mail.nwpu.edu.cn (H.C.); wangmingzhongsuper@163.com (M.W.); 554822887@qq.com (L.X.)
*Correspondence: xpzhao@nwpu.edu.cn (X.Z.)

**Abstract**

On the basis of the idea that the injecting energy will improve the conditions for the formation of Cooper pairs, a smart meta-superconductor (SMSC) was prepared by doping inhomogeneous phase of luminescent nanocomposite Y$_2$O$_3$:Eu$^{3+}$/Ag, which has the strong luminescence characteristic, in MgB$_2$ to improve the superconducting transition temperature ($T_C$) of the MgB$_2$-based superconductor. Two types of Y$_2$O$_3$:Eu$^{3+}$/Ag with different sizes were prepared and marked as m-Y$_2$O$_3$:Eu$^{3+}$/Ag and n-Y$_2$O$_3$:Eu$^{3+}$/Ag. MgB$_2$ SMSC was prepared through an ex situ process. Results show that when the inhomogeneous phase content was fixed at 2.0 wt.%, the $T_C$ of MgB$_2$ SMSC increased initially then decreased with the increase in the Ag content in the dopant. When the Ag content accounted for 5 wt.% of the inhomogeneous phase weight, the $T_C$ of MgB$_2$ SMSC was 37.2–38.0 K, which was similar to that of pure MgB$_2$. Meanwhile, the $T_C$ of MgB$_2$ SMSC doped with n-Y$_2$O$_3$:Eu$^{3+}$/Ag increased initially then decreased basically with the increase in the content of n-Y$_2$O$_3$:Eu$^{3+}$/Ag, in which Ag accounted for 5 wt.% of the inhomogeneous phase. The $T_C$ of MgB$_2$ SMSC doped with 0.5 wt.% n-Y$_2$O$_3$:Eu$^{3+}$/Ag was 37.6–38.4 K, which was 0.4 K higher than that of pure MgB$_2$. It is thought that the doping inhomogeneous phase of luminescent nanocomposite into the superconductor is a new means to improve the $T_C$ of SMSC.

**Introduction**

Improving the superconducting critical transition temperature of materials is an important scientific and technical problem in condensed matter physics and materials science. Recently, Fausti et al. [1] used mid-infrared femtosecond pulses to transform non-superconducting La$_{1.675}$Eu$_{0.2}$Sr$_{0.125}$CuO$_4$ into a transient 3D superconductor. A similar method was also applied to investigate YBa$_2$Cu$_3$O$_{6.5}$ [2] and K$_3$C$_{60}$ [3,4], and good experimental results were achieved. Ye et al.

have reported the observation of field-induced superconductivity of ZrNCl and $MoS_2$ by quasi-continuous electrostatic carrier doping achieved by combining liquid and solid gating [5,6]. Drozdov et al. [7] reported conventional superconductivity at 203 K under high pressure in a sulfur hydride system. Adu et al. [8] increased the $T_C$ of commercial "dirty" $MgB_2$ by conducting non-substitutional hole-doping of the $MgB_2$ structure using minute, single-wall carbon nanotube inclusions. In accordance with homogeneous system theory [9], Smolyaninov et al. [10-12] stated that a superconducting metamaterial with an effective dielectric response function that is less and approximately equal to zero may exhibit high $T_C$, and they verified this theory in their subsequent experiments. Recently, Cao et al. [13,14] investigated correlated insulator behavior at half-filling in magic-angle graphene superlattices and reported the realization of intrinsic unconventional superconductivity in a 2D superlattice created by stacking two sheets of graphene that are twisted relative to each other at a small angle. Another important method for studying superconductivity is the topological superconductors [15-20], which have attracted great attention in condensed matter physics. However, obtaining a practical superconductor with high $T_C$ remains difficult.

The superconductivity of $MgB_2$ was discovered in 2001 [21]. $MgB_2$ is a promising material with large-scale applications because of its excellent superconducting properties and simple crystal structure [22-27]. Considering that the $T_C$ of $MgB_2$ is close to the McMillan temperature limit [28,29], developing an effective experimental method to improve the $T_C$ of $MgB_2$ is beneficial to its practical application and to the understanding of the superconducting mechanism. Chemical doping is a simple, effective, commonly used method to change the $T_C$ of superconducting materials. However, many experimental results have confirmed that conventional chemical doping decreases the $T_C$ of $MgB_2$ [30-36]. To date, no effective method has been developed to improve the $T_C$ of $MgB_2$. The use of metamaterial structures to achieve special properties is an important method developed in recent decades [37-41], and it provides a new approach to improve the $T_C$ of superconducting materials.

On the basis of metamaterials, our group investigated the effects of ZnO electroluminescent (EL) material doping on the superconductivity of BSCCO in 2007 and attempted to change the $T_C$ of this superconductor [42]. Meanwhile, it is proposed that the combination of chemical doping and EL excitement, that is, doping EL materials in superconducting materials to form a meta structure, may be an effective method to improve the $T_C$ of superconductors [43]. On the basis of these results, a smart meta-superconductor (SMSC) model for improving the $T_C$ of materials has been proposed recently. In the model, the inhomogeneous phase is used to inject energy through its EL under the external field to strengthen the Cooper pairs, thereby achieving the purpose of changing the $T_C$. Zhang et al. [44] prepared $MgB_2$ doped with $Y_2O_3:Eu^{3+}$ particles through an in situ process. Tao et al. [45] prepared $MgB_2$ doped with $Y_2O_3:Eu^{3+}$ nanorods with different EL intensities through an ex situ process. Their results indicated that doping EL materials is favorable for the improvement of $T_C$ compared with conventional doping, which always reduces the superconducting transition temperature of the sample. In addition, similar experimental results were obtained by replacing $Y_2O_3$:$Eu^{3+}$ with $Y_3VO_4$:$Eu^{3+}$ flakes [46]. Meanwhlie, results also indicated that the $T_C$ can be changed by adjusting the $Y_2O_3$:$Eu^{3+}$ concentration and EL exciting current [47].

In this paper, a smart meta-superconductor was prepared by doping inhomogeneous phase of luminescent nanocomposite $Y_2O_3$:$Eu^{3+}$/Ag, which has the strong luminescence characteristic, in $MgB_2$ to improve the $T_C$ of the $MgB_2$-based superconductor. This inhomogeneous phase of nanocomposite illuminator $Y_2O_3$:$Eu^{3+}$/Ag was prepared by our group, its EL intensity is three times higher than that of $Y_2O_3$:$Eu^{3+}$ due to the composite illumination of electroluminescence and photoluminescence [48]. Two kinds of nanocomposite illuminator $Y_2O_3$:$Eu^{3+}$/Ag with different sizes, namely, micro $Y_2O_3$:$Eu^{3+}$/Ag (m-$Y_2O_3$:$Eu^{3+}$/Ag) and nano $Y_2O_3$:$Eu^{3+}$/Ag flakes (n-$Y_2O_3$:$Eu^{3+}$/Ag), are prepared. $MgB_2$ doped with a nanocomposite illuminator with high luminous intensity is prepared through an ex situ process [49]. The $T_C$ of the $MgB_2$-based superconductor is investigated by changing the Ag content in the inhomogeneous phase, the sizes of the inhomogeneous phase, and the doping concentration. The results indicate that the $T_C$ of $MgB_2$ doped with 0.5 wt.% n-$Y_2O_3$:$Eu^{3+}$/Ag is 37.6–38.4 K, which is 0.4 K higher than that of pure $MgB_2$.

**Experiment**

**1. Preparation of m-$Y_2O_3$:$Eu^{3+}$/Ag and n-$Y_2O_3$:$Eu^{3+}$/Ag**

$Y_2O_3$ (0.153 g) and $Eu_2O_3$ (0.012 g) were weighed and dissolved in a beaker with excess concentrated hydrochloric acid and subsequently heated and dried at 70 °C for 2 h to obtain a white precursor. One of the precursor was dissolved in 4 mL of deionized water to form a solution, and ammonium oxalate was added to it dropwise. The solution was subsequently stirred vigorously at 2 °C in a temperature-controlled water bath. A certain amount of $AgNO_3$ was added to the solution after been stirred for 30 min. After another 30 min of stirring, the pH value of the solution was adjusted to 9–10 by adding NaOH. The final solution, designated as solution A, was obtained after another 30 min of stirring. Another precursor was also prepared into solution with 24 mL benzyl alcohol. Octylamine (4 mL) was added dropwise to the solution, which was subsequently stirred for 1 h. Afterward, a certain amount of $AgNO_3$ was added to the solution. After stirring for another hour, another solution was obtained and designated as solution B. Solutions A and B were then transferred to two reaction kettles, respectively. A hydrothermal reaction occurred at 160 °C for 24 h. The products were washed several times with deionized water and absolute ethanol and sintered at 800 °C for 2 h to form $Y_2O_3$:$Eu^{3+}$/AgCl. After illumination, the $Y_2O_3$:$Eu^{3+}$/AgCl transformed into two kinds of luminescent $Y_2O_3$:$Eu^{3+}$/Ag nanocomposite with different sizes and a certain amount of

Ag. The two luminescent nanocomposite materials were designated as m-$Y_2O_3$:$Eu^{3+}$/Ag and n-$Y_2O_3$:$Eu^{3+}$/Ag. $Y_2O_3$:$Eu^{3+}$/Ag with different Ag contents was prepared by changing the $AgNO_3$ content. Meanwhile, similar method was applied to synthesize $Y_2O_3$ and $Y_2O_3$:$Sm^{3+}$.

## 2. Preparation of $MgB_2$–based SMSC

At a certain ratio, commercial $MgB_2$ powder and the luminescent nanocomposite $Y_2O_3$:$Eu^{3+}$/Ag were weighed and prepared into an alcohol solution. The two suspensions were sonicated for 20 min, then the dopant was added dropwise to $MgB_2$. After sonication for more than 20 min, the mixed solution was transferred to a culture dish. Subsequently, the culture dish was placed in a vacuum oven at 60 °C for 4 h to yield a black powder. The powder was pressed into a tablet and placed in a small tantalum container, which was annealed at 800 °C for 2 h in high-purity argon atmosphere. The $MgB_2$-based superconductor doped with luminescent nanocomposite materials of different sizes and Ag contents was synthesized to investigative the $T_C$ of SMSC.

## Results and discussion

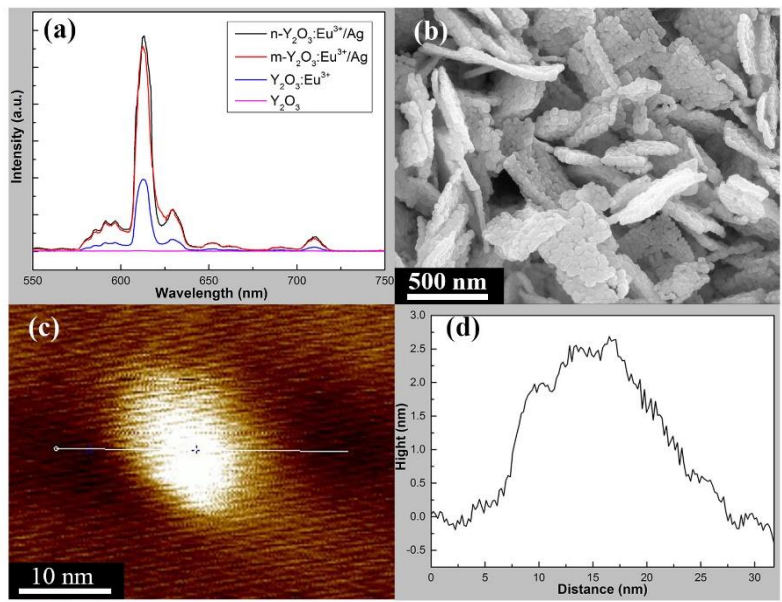

**Figure 1.** (a) EL spectra of $Y_2O_3$, $Y_2O_3$:$Eu^{3+}$, m-$Y_2O_3$:$Eu^{3+}$/Ag, and n-$Y_2O_3$:$Eu^{3+}$/Ag;
(b) SEM image of m-$Y_2O_3$:$Eu^{3+}$/Ag; and (c-d) AFM image of n-$Y_2O_3$:$Eu^{3+}$/Ag.

Figure 1a shows the EL spectra of $Y_2O_3$, $Y_2O_3$:$Eu^{3+}$, m-$Y_2O_3$:$Eu^{3+}$/Ag, and n-$Y_2O_3$:$Eu^{3+}$/Ag. The Ag content of the luminescent nanocomposite materials was 5.0 wt.%. It shows that $Y_2O_3$ is a non-EL material and becomes a kind of EL material after the addition of a small amount of Eu element. The results also indicate that the EL intensity of the luminescent m-$Y_2O_3$:$Eu^{3+}$/Ag nanocomposite and n-$Y_2O_3$:$Eu^{3+}$/Ag is remarkably improved primarily due to the composite luminescence of the electroluminescence of $Eu^{3+}$ centric and the surface plasma-enhanced photoluminescence of Ag. Among the four dopants, n-$Y_2O_3$:$Eu^{3+}$/Ag had the highest EL intensity. Figure 1b shows the SEM image of m-$Y_2O_3$:$Eu^{3+}$/Ag. The surface size and thickness of the m-$Y_2O_3$:$Eu^{3+}$/Ag flake are approximately 300 nm and 30 nm, respectively. Figs. 1c–1d show AFM images of n-$Y_2O_3$:$Eu^{3+}$/Ag. The surface size of n-$Y_2O_3$:$Eu^{3+}$/Ag was 20 nm, and its thickness was approximately 2.5 nm, which is much smaller than that of m-$Y_2O_3$:$Eu^{3+}$/Ag.

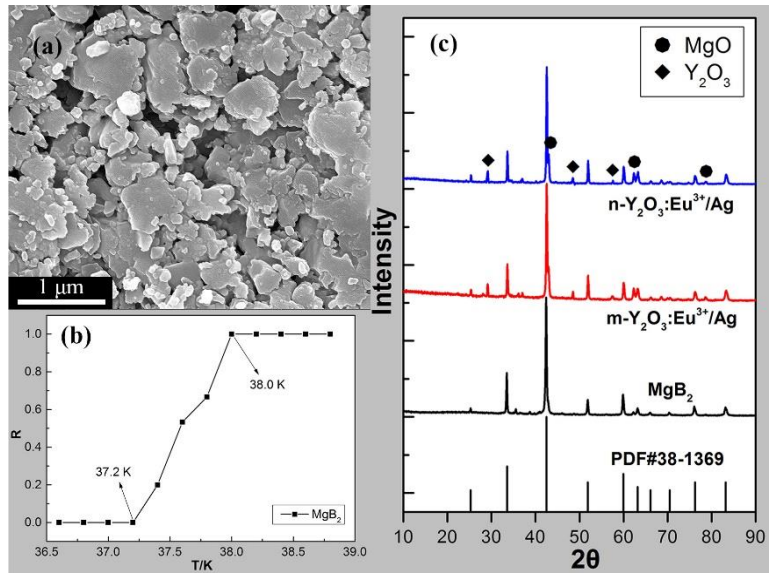

**Figure. 2** (a) SEM image and (b) normalized temperature-dependent resistivity (*R*–*T*) curve of pure $MgB_2$; (c) XRD spectra of pure $MgB_2$ and $MgB_2$ doped with m-$Y_2O_3$:$Eu^{3+}$/Ag and n-$Y_2O_3$:$Eu^{3+}$/Ag.

Figure 2a shows the SEM image of pure $MgB_2$. The size of the $MgB_2$ particle was approximately 0.1–1 μm. The $T_C$ of the samples was determined based on the *R*–*T* curve, which was measured using a four-probe method in a liquid helium cryogenic system developed by the Advanced Research Systems Company. Figure 2b shows the normalized *R*–*T* curve of pure $MgB_2$ and indicates that the onset temperature ($T_c^{on}$) and offset temperature ($T_c^{off}$) [50,51] of pure $MgB_2$ were 38.0 and 37.2 K, respectively. The superconducting transition width (*ΔT*) of pure $MgB_2$ was 0.8 K. Figure 2c shows the XRD spectra of pure $MgB_2$ and partially doped samples, in which the standard card of $MgB_2$ (PDF#38-1369) is demonstrated using black vertical lines. The results showed that the XRD spectrum of pure $MgB_2$ (black curve) matched the standard card of $MgB_2$ well, except for the inevitable small amount of the MgO phase [52-55]. The red and blue curves represent the XRD spectra of $MgB_2$ doped with 2.0 wt.% m-$Y_2O_3$:$Eu^{3+}$/Ag and 2.0 wt.% n-$Y_2O_3$:$Eu^{3+}$/Ag, respectively. The Ag content was 5.0 wt.% of the dopant weight. The main phase of the doped samples was $MgB_2$. Moreover, apart from a small amount of the MgO phase, the $Y_2O_3$ phase was also found in the XRD spectra of the doped samples. The XRD spectra of the other doped samples were similar.

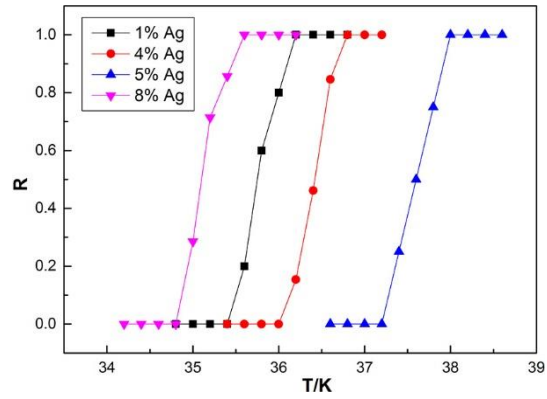

**Figure 3.** Normalized *R*–*T* curves of MgB$_2$ doped with 2.0 wt.% m-Y$_2$O$_3$:Eu$^{3+}$/Ag with different Ag contents.

Figure 3 shows the normalized *R*–*T* curves of MgB$_2$ doped with m-Y$_2$O$_3$:Eu$^{3+}$/Ag with different Ag contents. On the basis of the results of our previous study [45,46], the content of m-Y$_2$O$_3$:Eu$^{3+}$/Ag in the four samples was fixed at 2.0 wt.%. The Ag contents of m-Y$_2$O$_3$:Eu$^{3+}$/Ag in the four samples were 1 wt.%, 4 wt.%, 5 wt.%, and 8 wt.%, as shown in the figure, and their $T_C$ values were 34.8–35.6, 36.0–36.8, 37.2–38.0, and 34.8–35.6 K, respectively. The $T_C$ of the doped samples initially increased then decreased with the increase in Ag content. Meanwhile, the corresponding doped sample had the highest $T_C$ when the concentration of m-Y$_2$O$_3$:Eu$^{3+}$/Ag was fixed at 2.0 wt.% and the Ag content of m-Y$_2$O$_3$:Eu$^{3+}$/Ag was 5 wt.%, which is equal to that of pure MgB$_2$. The experimental results are similar to those of our previous studies, that is, doping EL materials may improve $T_C$ in several cases compared with conventional doping, which always reduces the $T_C$ of the sample. As a dopant, m-Y$_2$O$_3$:Eu$^{3+}$/Ag exerts an impurity effect that decreases $T_C$. Meanwhile, as an EL material, m-Y$_2$O$_3$:Eu$^{3+}$/Ag exerts an EL exciting effect that increases $T_C$ [45,46]. An obvious competitive relationship exists between the impurity effect and the EL exciting effect. The final $T_C$ of the samples increased when the EL exciting effect was fully utilized and the impurity effect was minimized.

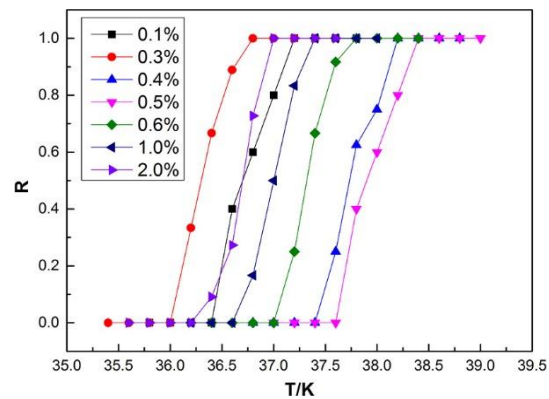

**Figure 4.** Normalized *R*–*T* curves of MgB$_2$ doped with n-Y$_2$O$_3$:Eu$^{3+}$/Ag.

Figure 4 shows the normalized $R$–$T$ curves of $MgB_2$ doped with 0.1–2.0 wt.% n-$Y_2O_3$:$Eu^{3+}$/Ag. Ag concentration was fixed at 5.0 wt.% of the weight of n-$Y_2O_3$:$Eu^{3+}$/Ag. It can be seen that $T_C$ of $MgB_2$ doped with n-$Y_2O_3$:$Eu^{3+}$/Ag initially decreased, increased, then decreased again with the increase in doping concentration. A too low or too high doping concentration reduces $T_C$, which is similar to the finding of our previous study. When the doping concentration was in a low range, $T_C$ decreased with the increase in doping concentration due to the dominance of the impurity effect of the dopant, which is similar to the results of conventional doping. The EL exciting effect of the dopant dominated with the further increase in doping concentration, resulting in the increase in $T_C$. The samples doped with 0.5 wt.% n-$Y_2O_3$:$Eu^{3+}$/Ag had the highest $T_C$ of 37.6–38.4 K, which is 0.4 K higher than that of pure $MgB_2$. However, the impurity effect of the dopant dominated when the doping concentration increased to a high range, which led to a low $T_C$. These results indicate that doping luminescent nanocomposite materials effectively adjusts and improves $T_C$ at an appropriate doping concentration.

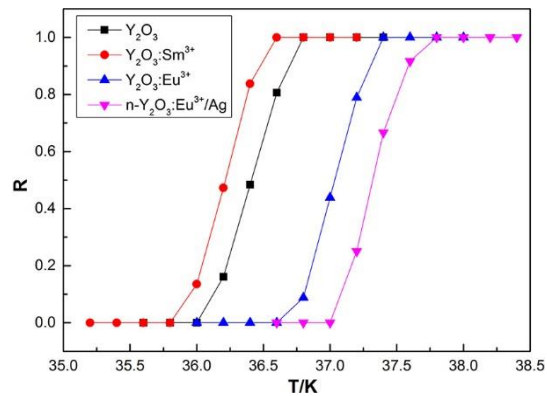

**Figure 5.** Normalized $R$–$T$ curves of $MgB_2$ doped with $Y_2O_3$, $Y_2O_3$:$Sm^{3+}$, $Y_2O_3$:$Eu^{3+}$, and n-$Y_2O_3$:$Eu^{3+}$/Ag.

$MgB_2$ doped with non-EL materials $Y_2O_3$ and $Y_2O_3$:$Sm^{3+}$ were synthesized to prove the conclusions above. Figure 5 shows the normalized $R$–$T$ curves of $MgB_2$ doped with $Y_2O_3$, $Y_2O_3$:$Sm^{3+}$, $Y_2O_3$:$Eu^{3+}$, and n-$Y_2O_3$:$Eu^{3+}$/Ag. The doping concentration was fixed at 0.6 wt.%, and the Ag content in n-$Y_2O_3$:$Eu^{3+}$/Ag was 5.0 wt.%. Results indicated that $T_C$ of $MgB_2$ doped with non-EL materials $Y_2O_3$ or $Y_2O_3$:$Sm^{3+}$ was much lower than that of pure $MgB_2$, which is different from $MgB_2$ doped with EL materials at the same concentration. $MgB_2$ doped with $Y_2O_3$:$Eu^{3+}$ increased to 36.6–37.4 K due to the EL exciting effect. Meanwhile, $MgB_2$ doped with the luminescent n-$Y_2O_3$:$Eu^{3+}$/Ag nanocomposite further increased to 37.0–37.8 K. The results show that doping EL materials facilitates an increase in $T_C$ in several cases compared with conventional doping, which always reduces the $T_C$ of the sample. Meanwhile, luminescent $Y_2O_3$:$Eu^{3+}$/Ag nanocomposite materials increase the $T_C$ of $MgB_2$ due to the strong EL intensity.

The results in Figure 4 show that the optimum concentration of n-$Y_2O_3$:$Eu^{3+}$/Ag is 0.5 wt.%,

which is lower than the value in our previous study [44-46] due to the small size of n-$Y_2O_3$:$Eu^{3+}$/Ag. The disadvantages caused by the impurity effect can be reduced if luminescent nanocomposite materials have a small size and are relatively evenly distributed in the sample. Moreover, the $\Delta T$ of commercial $MgB_2$ in our previous study [46] was too large to accurately determine the influence of the inhomogeneous phase on $T_C$. In the current study, a new kind of commercial $MgB_2$ with a small $\Delta T$ of 0.8 K was used, and we obtained a similar conclusion, which further proves the effectiveness of this method.

**Conclusion**

On the basis of the idea that the injecting energy will improve the conditions for the formation of Cooper pairs, a smart meta-superconductor was prepared by doping inhomogeneous phase of luminescent nanocomposite $Y_2O_3$:$Eu^{3+}$/Ag in $MgB_2$ to improve the $T_C$ of the $MgB_2$-based superconductor. Two types of $Y_2O_3$:$Eu^{3+}$/Ag with different sizes were prepared and marked as m-$Y_2O_3$:$Eu^{3+}$/Ag and n-$Y_2O_3$:$Eu^{3+}$/Ag. $MgB_2$ SMSC was prepared through an ex situ process. SEM and AFM images indicated that the surface size and thickness of m-$Y_2O_3$:$Eu^{3+}$/Ag are approximately 300 nm and 30 nm, which are 20 nm and 2.5 nm for n-$Y_2O_3$:$Eu^{3+}$/Ag. The EL spectra showed that the EL intensity of luminescent nanocomposite $Y_2O_3$:$Eu^{3+}$/Ag is three times higher than that of $Y_2O_3$:$Eu^{3+}$. The $T_C$ of $MgB_2$ SMSC was determined based on the measured $R$–$T$ curve by using the four-probe method in a liquid helium cryogenic system. Results show that when the inhomogeneous phase content was fixed at 2.0 wt.%, the $T_C$ of $MgB_2$ SMSC initially increased then decreased with the increase in the Ag content in the dopant. When the Ag content accounted for 5 wt.% of the inhomogeneous phase weight, the $T_C$ of $MgB_2$ SMSC was 37.2–38.0 K, which was similar to that of pure $MgB_2$. Meanwhile, the $T_C$ of $MgB_2$ SMSC doped with n-$Y_2O_3$:$Eu^{3+}$/Ag increased initially then decreased basically with the increase in the content of n-$Y_2O_3$:$Eu^{3+}$/Ag, in which Ag accounted for 5 wt.% of the inhomogeneous phase. The $T_C$ of $MgB_2$ SMSC doped with 0.5 wt.% n-$Y_2O_3$:$Eu^{3+}$/Ag was 37.6–38.4 K, which was 0.4 K higher than that of pure $MgB_2$. It is thought that the doping inhomogeneous phase of luminescent nanocomposite into the superconductor is a new means to improve the $T_C$ of SMSC.

**Funding:** This work was supported by the National Natural Science Foundation of China for Distinguished Young Scholar under Grant No. 50025207.